\documentclass[conference]{IEEEtran}
\IEEEoverridecommandlockouts

\usepackage{cite}
\usepackage{amsmath,amssymb,amsfonts}
\usepackage{algorithm} %
\usepackage{graphicx}
\usepackage{textcomp}
\usepackage{xcolor}

\usepackage{array}
\usepackage[caption=false,font=normalsize,labelfont=sf,textfont=sf]{subfig}
\usepackage{stfloats}
\usepackage{url}
\usepackage{verbatim}
\usepackage{enumitem}
\usepackage{caption}
\usepackage[normalem]{ulem}
\usepackage{multirow}
\usepackage{multicol}
\usepackage{amsmath,algpseudocode}%

\def\BibTeX{{\rm B\kern-.05em{\sc i\kern-.025em b}\kern-.08em
    T\kern-.1667em\lower.7ex\hbox{E}\kern-.125emX}}

\algdef{SE}[EVENT]{Event}{EndEvent}[1]{\textbf{upon event}\ #1\ \algorithmicdo}{\algorithmicend\ \textbf{event}}%
\algtext*{EndEvent}

\begin{document}

\title{Hardware-based Scheduler Implementation for Dynamic Workloads on Heterogeneous SoCs \\
{\footnotesize
\thanks{This material is based on research sponsored by Air Force Research Laboratory (AFRL) and Defense Advanced Research Projects Agency (DARPA) under agreement number FA8650-18-2-7860. The U.S. Government is authorized to reproduce and distribute reprints for Governmental purposes notwithstanding any copyright notation thereon. The views and conclusion contained herein are those of the authors and should not be interpreted as necessarily representing the official policies or endorsements, either expressed or implied, of Air Force Research Laboratory (AFRL) and Defence Advanced Research Projects Agency (DARPA) or the U.S. Government.}}}

\author{\IEEEauthorblockN{Alexander Fusco, Sahil Hassan, Joshua Mack, Ali Akoglu}
\IEEEauthorblockA{\textit{Electrical and Computer Engineering}, \textit{The University of Arizona}\\
Tucson, AZ, USA \\
\{afusco1,sahilhassan,jmack2545,akoglu\}@email.arizona.edu}
}

\maketitle

\begin{abstract}

Non-uniform performance and power consumption across the processing elements (PEs) of heterogeneous SoCs increase the computation complexity of the task scheduling problem compared to homogeneous architectures. 
Latency of a software-based scheduler with the increased heterogeneity level in terms of number and types of PEs  creates the necessity of deploying a scheduler as an overlay processor in hardware to be able to make scheduling decisions rapidly and enable deployment of real-life applications on heterogeneous SoCs.
In this study we present the design trade-offs involved for implementing and deploying the runtime variant of the heterogeneous earliest finish time algorithm (HEFT\textsubscript{RT}) on the FPGA. We conduct performance evaluations on a SoC configuration emulated over the Xilinx Zynq ZCU102 platform. In a runtime environment we demonstrate hardware-based HEFT\textsubscript{RT}'s ability to make scheduling decisions with 9.144 ns latency on average, process 26.7\% more tasks per second compared to its software counterpart, and reduce the scheduling latency by up to a factor of 183$\times$ based on workloads composed of mixture of dynamically arriving real-life signal processing applications.

\end{abstract}


\section{Introduction}
Heterogeneous system on chip (SoC) computing platforms composed of a pool of general-purpose processors and specialized processors offer the ability to pair each execution phase of a given application with a compute unit that matches the processing needs of that phase. Hence, they offer a large potential for performance and energy-efficiency relative to homogeneous counterparts \cite{cdsc}. However, from task scheduling point of view, heterogeneous SoCs pose additional challenges as they provide a variety of compute resources giving tasks varying execution times that create a larger scheduling search space to explore. When building runtime systems for heterogeneous SoCs, it is difficult to design schedulers that are capable of thoroughly exploring the space of possible schedules while still returning effective scheduling decisions within a reasonable amount of time. Effective utilization of system resources while supporting execution of multiple interleaving applications requires a dynamic scheduling policy that considers task dependencies within an application and demands on the same processing element (PE) by multiple applications. When systems are highly heterogeneous in terms of both the number and types of PEs, and when a runtime system must deal with numerous applications co-existing on the system at once, the latency of software-based schedulers can become a major bottleneck to total system throughput. As a result, there is a strong case to be made that scheduling algorithms themselves should be subject to hardware-based acceleration to enable utilization of heterogeneous SoC platforms for practical deployment under scenarios where multiple applications co-exist in domains such as autonomous vehicles and communication systems that require low Size, Weight and Power (SWaP) solutions.

In this study, we utilize the HEFT\textsubscript{RT}~\cite{Mack22} algorithm, a runtime variant of the Heterogeneous Earliest Finish Time (HEFT)~\cite{Topcuoglu02} scheduling heuristic, implement it in hardware as an overlay processor for an SoC composed of ARM CPU cores and FFT accelerator that is emulated on the Xilinx Zynq ZCU102 board. 
We analyze the design trade-offs involved for implementing and deploying the HEFT\textsubscript{RT} on the FPGA. We perform this analysis with realistic hardware configurations coupled with complex signal-processing workloads that give confidence that the results seen here will generalize across other applications.  
We show that for a runtime system with ready queue size of up to 5 tasks, the software based scheduler should be preferred. 
Beyond this point we observe that the data transfer overhead to and from the hardware-based scheduler is compensated by the reduced scheduling latency.
As the ready queue size increases, the gap in scheduling overhead between hardware-based HEFT\textsubscript{RT} and software-based HEFT\textsubscript{RT} grows.
Previous literature has found that hardware-accelerated scheduling is beneficial in homogeneous systems. 
In this study, we demonstrate the ability to accelerate a high quality, heterogeneity-aware scheduling heuristic on hardware in a manner that demonstrably reduces scheduling overhead and consequently improves total system performance.
We show that hardware-based HEFT\textsubscript{RT} is capable of making up to 183$\times$ faster scheduling decisions across the ready queue sizes used in our evaluations. 
Before presenting our approach to implementing HEFT\textsubscript{RT} in hardware, we start by exploring related work in the area of task scheduling with emphasis on hardware based dynamic task scheduling.

\section{Related Work} \label{sec:literature}

Previous hardware schedulers have implemented deadline-based scheduling policies~\cite{fpga_edf, hrhs, hw_pfair, red}, with most of them relying on earliest deadline first (EDF) for real time systems that has been shown to minimize the maximum lateness and meet deadlines when a feasible solution exists~\cite{rt_scheduling}.

The study by Tang and Bergman~\cite{fpga_edf} discuss types of task queues and their benefits across scheduling policies such as EDF, deadline based least slack time, and multilevel feedback queue using static priorities. Derafshi et al. \cite{hrhs} expands upon Tang and Bergman's study~\cite{fpga_edf} by targeting a homogeneous multi-core processor, and moving more task dependency management to hardware. The hardware scheduler by Gupta et al.~\cite{hw_pfair} also targets multi-core homogeneous processors, and implements Pfair~\cite{pfair}, a scheduling policy that uses the expected execution time to estimate when a task will finish in relation to its deadline. Kohútka et al.\cite{red} implement a hardware scheduler that aims to improve on the EDF policy by discarding tasks when the workload becomes unfeasible.

These hardware schedulers do not consider runtime constraints regarding how the task-to-PE-mapping decision for a given application may affect the execution time of other dynamically arriving applications, and they are limited to targeting single-core~\cite{fpga_edf, red} or multi-core homogeneous architectures~\cite{hrhs, hw_pfair}. To the best of our knowledge, the only previous work on hardware-based scheduler for heterogeneous architectures is presented by Aliyev et al.~\cite{hw_heft}, which implements the Heterogeneous Earliest Finish Time~\cite{Topcuoglu02}. This implementation, unlike EDF based schedulers, is not suitable for runtime execution as it requires parsing the full directed acyclic graph (DAG) of an application, and can only perform scheduling on one application at a time. Among all the hardware schedulers, the implementation by Derafshi et al.\cite{hrhs} is the only one that integrates their scheduler into a runtime environment, where they perform emulation based analysis using artificial tasks. 
None of these studies present analysis on the quality of schedules generated by the hardware scheduler with dynamic workloads. 
We perform such an analysis by implementing HEFT\textsubscript{RT} in hardware and integrating with a runtime environment.

\section{Background} \label{sec:background}

\subsection{Runtime Environment}\label{sec:emulation}
In this study, we utilize the Compiler Integrated Extensible DSSoC Runtime (CEDR) ecosystem~\cite{MackTECS22} as it enables compilation and development of user applications, evaluation of resource management strategies, and validation of heterogeneous hardware configurations in a unified framework. This runtime system operates as a background \textit{Daemon Process} in the Linux user space and allows end users to interact with the heterogeneous architecture by launching a workload composed of any number and combination of distinct applications with user specified arrival rates. 
CEDR launches worker threads, each tasked with managing a particular PE on the heterogeneous system in terms of sending data to its respective PE, executing its assigned tasks, and receiving their output. Meanwhile, the CEDR management thread parses the incoming application DAGs, monitors the state of tasks executing on each PE, maintains a ready queue with tasks whose dependencies are resolved, and estimates when a PE will become available. 
A mapping event occurs when a new task arrives in the ready queue, which invokes the scheduler. The tasks that are in the ready queue and the estimated availability times of the PEs are then passed to the "Scheduler" at each mapping event.

\subsection{Runtime Variant: HEFT\textsubscript{RT}}

Heterogeneous Earliest Finish Time (HEFT)~\cite{Topcuoglu02} is a scheduling heuristic that balances runtime complexity and quality of generated schedules, and has been shown to generate competitive schedules to this date~\cite{maurya2018benchmarking}.
List-scheduling algorithms, such as HEFT, are typically used as purely static schedulers as they operate on DAGs with known task dependencies and application structures. 
The study by Mack et al.~\cite{Mack22} introduces a methodology that lifts the barriers of list schedulers from being deployed in a runtime environment through a series of transformations. They apply this methodology and develop the runtime variant HEFT\textsubscript{RT}. 
At each mapping event, HEFT\textsubscript{RT} parses the ready queue, PE state, earliest availability time collected from the runtime. It sorts all tasks in the ready queue using a separate priority queue in order of highest average runtime across all PEs. As tasks are dequeued in priority order one by one, for each task, the PE that delivers the earliest finish time is selected. This selection along with the updated availability time for that selected PE are finally sent to the runtime system. 
In the following section we discuss the proposed hardware implementation of this flow.

\section{HEFT\textsubscript{RT} Hardware Implementation}\label{sec:hardware}

\subsection{AXI Stream Queuing Interface}
\begin{figure}[t]
    \centering
    \includegraphics[width=0.80\linewidth, trim=3cm 0.5cm 3cm 0.5cm]{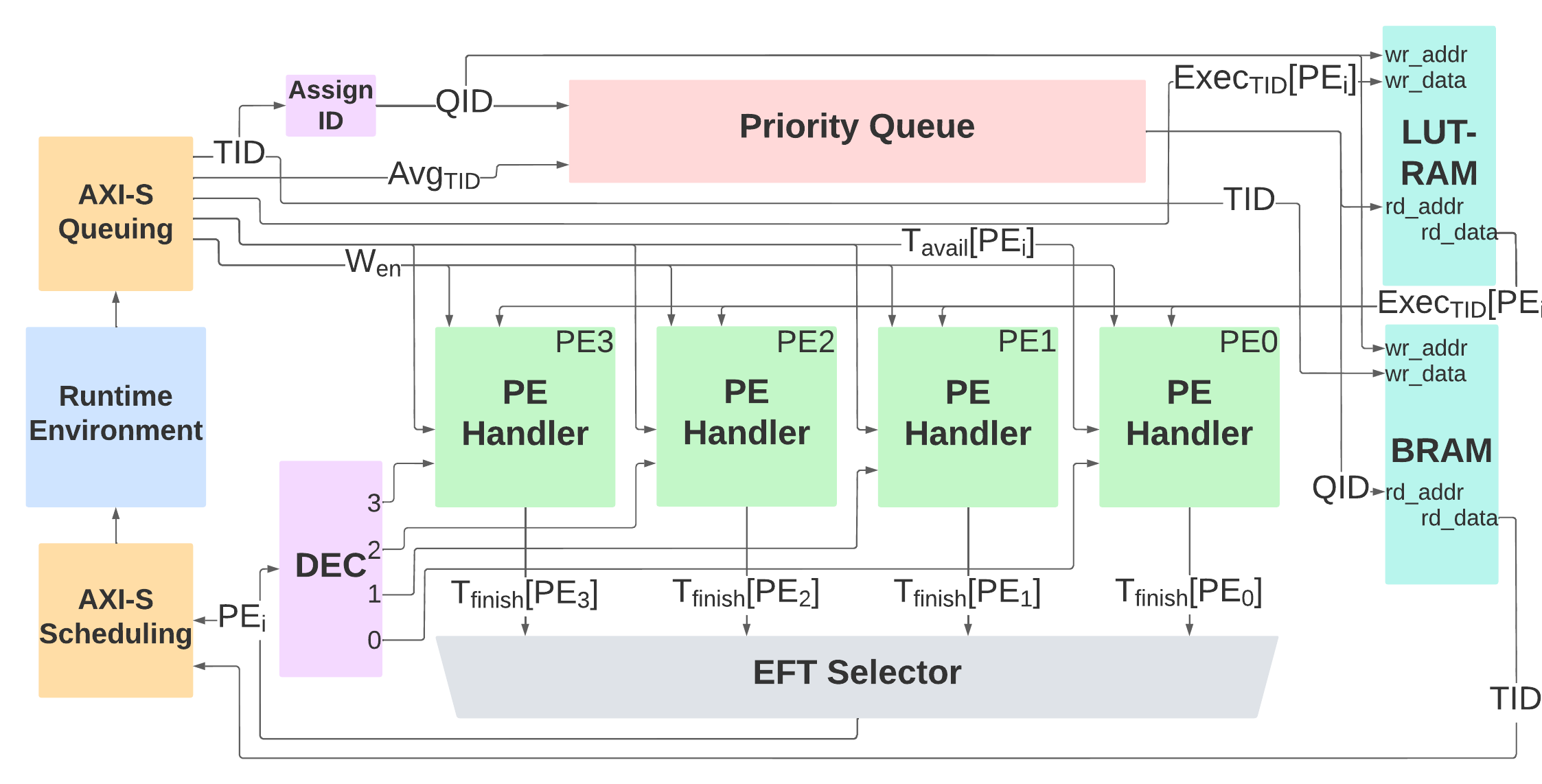}
    \caption{Architecture of hardware scheduler.}
    \label{fig:hw_scheduler}
    \vspace{-4pt}
\end{figure}

The implementation of the HEFT\textsubscript{RT} algorithm is illustrated in Figure \ref{fig:hw_scheduler}. For figure clarity, the definitions and expressions for related parameters are listed in Table~\ref{tab:parameters}. 
When interfacing the hardware based HEFT\textsubscript{RT} implementation with the runtime environment and the processing elements of the emulated SoC, we use direct memory access (DMA) blocks to move data between the host CPU and the processing elements via the AXI4-Stream protocol. 
At each mapping event, the CEDR sends the updated availability times for each PE ($T_{avail}$) to the scheduler, which are forwarded to its designated \emph{PE Handler}. This synchronizes the availability times with the runtime environment before processing the ready queue. For each task in the ready queue, the runtime environment feeds the task information that includes a unique task identifier ($TID$), average execution time across the PEs ($Avg_{TID}$), and execution time on each PE ($Exec_{TID}[PE_{i}]$), to the scheduler through an AXI-Stream (AXI-S). The task is assigned a counter-based queue identifier ($QID$) ranging from 0 to the depth of the priority queue ($D$). 
The mapping of $TID$ to $QID$ ensures that each task is identified within a fixed range.
The $QID$ and $Avg_{TID}$ are injected into the priority queue. While the $TID$ is written into the BRAM, the execution times ($Exec_{TID}[PE_{i}]$) are written into the LUT-RAM, both at addresses specified by the $QID$. It takes one cycle to insert the task information into the \emph{Priority Queue} and memory blocks once they are received from the runtime.

\begin{table}[!t]
\caption{\small{Hardware Scheduler Parameters\label{tab:parameters}}}
\label{tab:params}
\centering
{
\renewcommand{\arraystretch}{1.1}
\begin{tabular}{ l l }
\hline
Parameter & Meaning  \\
\hline
$P$ & Number of PEs \\
$D$ & Depth of \emph{Priority Queue} \\
$PE_{i}$ & Identifier of a PE where $i \in [0,P)$ \\
$TID$ & Task identifier \\
$QID$ & Queue identifier \\
$Avg_{TID}$ & Average execution time of task  across all PEs \\
$Exec_{TID}[PE_{i}]$ & Task execution time for a given PE \\
$T_{avail}$ & $T_{avail}[PE_{i}]$ \\
$T_{finish}$ & $T_{finish}[PE_{i}]$ \\
$W(Avg_{TID})$ & Bit width of $Avg_{TID}$ \\
$W(QID)$ & Bit width of $QID$, which is $\lceil log_{2}D \rceil$ \\
\hline
\end{tabular}
}
\vspace{-6pt}
\end{table}

\begin{figure}[t]
    \centering
    \includegraphics[width=0.95\linewidth, trim=0.6cm 0.5cm 0.5cm 0.5cm, clip=true]{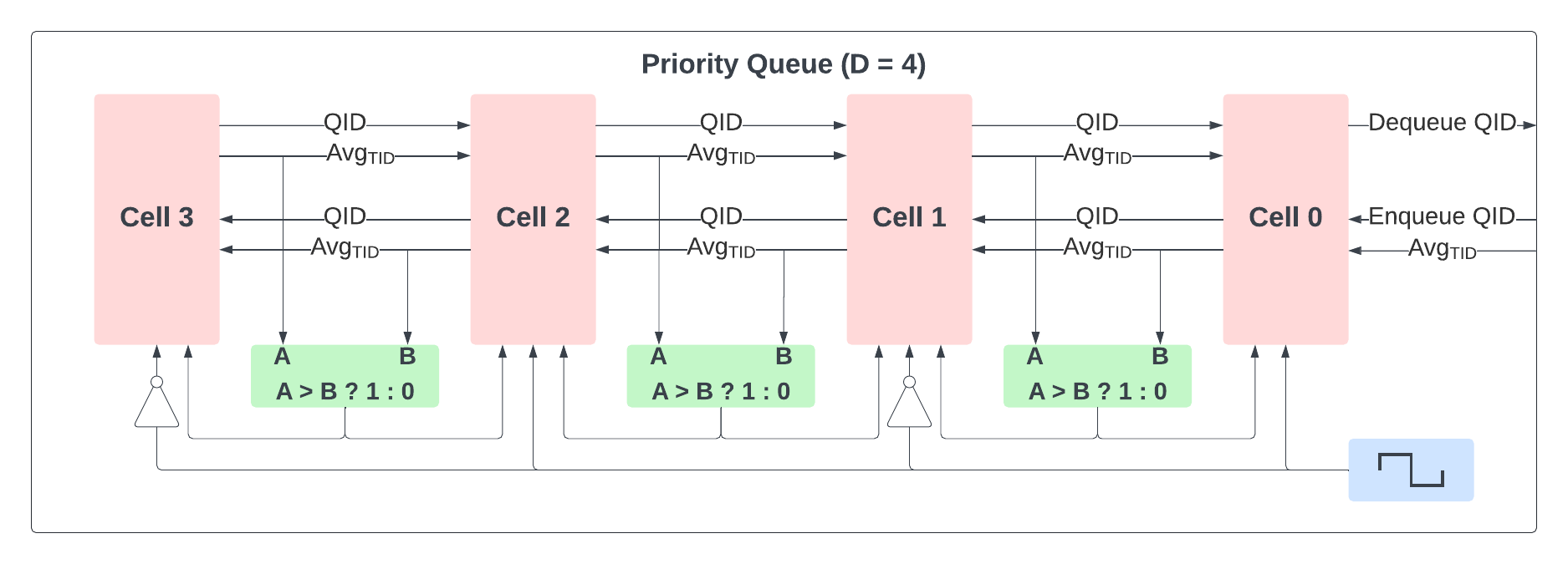}
    \caption{Architecture of shift register priority queue.}
    \label{fig:priority_queue}
    \vspace{-10pt}
\vspace{-0.2cm}
\end{figure}

\subsection{Priority Queue}
At each mapping event, the hardware scheduler sorts the tasks that are in the \emph{Priority Queue}. Each cell in the queue, as illustrated in Figure \ref{fig:priority_queue}, includes the task information in terms of $QID$ and $Avg_{TID}$, where average execution time represents the relative importance of each task.

The \emph{Priority Queue} is implemented as a shift register priority queue that has also been used in EDF based hardware schedulers \cite{fpga_edf, red}. Each cell in the queue swaps its task information with its neighbor when the condition of the comparator logic is satisfied. The comparison operations are carried out alternately among the odd and even indexed cells following the odd-even sorting transposition algorithm~\cite{odd_even_sort}. Therefore it takes two clock cycles to complete every comparison. When there are two cycles in a row without any swaps, the sorting process is terminated. Then the queue switches to right-shift register mode and starts dequeuing the cell entries until all entries have been read. 
In each clock cycle, the output $QID$ is used as an index to read out execution times for each PE from the LUT-RAM to be forwarded to their corresponding \emph{PE Handler} along with the $TID$ to be forwarded to the \emph{AXI-S Scheduling Interface}.

\subsection{Processing Element Handler}\label{sec:hardware_pehandler}
For each PE in the target hardware configuration there is a designated \emph{PE Handler} in the hardware scheduler. It receives the $T_{avail}$ from the AXI-S and stores this information in the \emph{availability time} register of the \emph{PE Handler}. The output is the finish time ($T_{finish}$), which is the sum of $Exec_{TID}[PE_{i}]$ and $T_{avail}[PE_{i}]$. Finish time represents the estimated time when that task will complete its execution on that PE. This finish time is written into the \emph{availability time} register only if that PE is selected by the \emph{EFT Selector}.

The \emph{Earliest Finish Time (EFT) Selector} receives the finish time for each PE as an input, and outputs an index corresponding to the index of the PE ($PE_{i}$) with the lowest finish time. A comparator tree is utilized to implement the selection process. The \emph{EFT Selector} feeds the index to each of the \emph{PE Handlers}. The handler of the selected PE then makes the calculated finish time the new availability time of that PE. During this update, the $TID$ and the $PE_{i}$ are sent to the runtime through the AXI-S Scheduling Interface.

\section{Experimental Setup}
\label{sec:setup}
All experiments are performed on the Xilinx Zynq Ultrascale+ ZCU102 MPSoC development board.
This MPSoC combines general purpose CPUs (4x Arm Cortex A53 processors) with programmable FPGA fabric. On the FPGA fabric, we add the Xilinx FFT IP accelerator for Fast Fourier Transform (FFT). We use four representative real-world applications part of the CEDR release~\cite{MackTECS22} that are Radar Correlator (RC), Temporal Interference Mitigation (TM), Pulse Doppler (PD), and WiFi TX (TX). Based on average observed latencies, RC and TM are \textit{low} latency applications, whereas relatively TX and PD are \textit{high} latency applications.  We refer to the \emph{"Frame"} as the input data size needed by a single instance of each of the applications. The \textit{low} workload consists of twenty \emph{Frames} for RC and TM for a 1280Kb per frame. We compose a \textit{high} latency workload consisting of ten instances of PD and TX applications each, with a \emph{Frame} size of 1037 Kb per frame.
We use the \emph{low}-latency workload for verifying the functionality of the hardware HEFT\textsubscript{RT} with respect to its software counterpart, as low latency workloads will not stress the scheduler and in turn trends will not be affected by the hardware based scheduler making faster scheduling decisions. The high latency workloads on the other hand will allow us to compare the performance of the two schedulers when the system becomes oversubscribed. We evaluate the performance of each scheduler in this study by stressing the scheduler's ability to cope with the volume of incoming jobs.
We establish this by setting a fixed "target" frame rate that defines the rate of application instances arriving in the runtime per second. 
Then, we monitor if the scheduler can complete jobs at that same rate or if system performance begins to suffer.
We refer to the empirical frame rate observed as the "achieved" frame rate, which indicates the number of applications that the runtime can process per second.
For this, we use 29 different injection rates, where the injection rate defines the amount of input data arriving into the runtime per unit time. 

We use three metrics for performance evaluation. The cumulative execution time of an application is the sum of execution times of its individual tasks, ignoring the overhead associated with scheduling them. Lower cumulative execution time indicates better scheduling decisions made and better exploitation of the available heterogeneity at hand. The scheduling overhead metric captures the time spent by the runtime in making scheduling decisions.
The application execution time is the difference between the end of the last task and the start of the first task of an application, including the overhead of all scheduling decisions in between. Lower execution times indicate the scheduler's capability to manage the workload efficiently.
To make the cumulative execution times and application execution times comparable across different runtime configurations, we normalize them with the number of applications (per application).
We sweep all input configurations, repeat each experiment twenty five times, and collect performance metrics listed in the table by averaging across the twenty five runs.
We then average these three metrics across twenty five repetitions.

\section{Results and Analysis}\label{sec:results}

\subsection{Hardware Analysis}

\begin{table}[t]
\caption{\small{Resource usage on Zynq ZCU102 FPGA along with the total including peripheral resources shown in Figure~\ref{fig:hw_scheduler}.}
}
\label{tab:components}
\centering
\resizebox{1.0\linewidth}{!}
{
\begin{tabular}{| c | c | c | c | c | c | c | c | c |}
\hline
Module & \multicolumn{2}{c|}{Logic LUTs} & \multicolumn{2}{c|}{LUT-RAM} & \multicolumn{2}{c|}{Registers} & \multicolumn{2}{c|}{BRAM} \\
\hline
Priority Queue & 18,632 & 6.8\% & 0 & 0\% & 13,433 & 2.5\% & 0 & 0\% \\
PE Handlers & 404 & 1.5\% & 0 & 0\% & 128 & 0.5\% & 0 & 0\% \\
EFT Selector & 48 & 0.02\% & 0 & 0\% & 0 & 0\% & 0 & 0\% \\
Total & 19,603 & 7.2\% & 1,280 & 0.9\% & 14,534 & 2.7\% & 0.5 & 0.05\% \\

\hline
\end{tabular}
}
\vspace{-4pt}
\end{table}

Table \ref{tab:components} shows resource usage of the HEFT\textsubscript{RT} scheduler on the Zynq ZCU102 for a configuration where the system is composed of 4 PEs, priority queue depth is 512, and bit-width for representing average execution time ($W(Avg_{TID})$) is set to 16. We choose 16 bits over 32 bits, since this bit-width provides sufficient dynamic range and reduces the amount of LUTs used by the  datapath. 
The design uses 7.15\% of the available LUTs and the majority of the LUTs and registers in the scheduler are consumed by the \emph{Priority Queue}. The amount of LUT-RAM reserved for storing the execution time information of each DAG node for each PE is a product of the number of PEs and the depth of the priority queue, since it stores the execution time for each task on each PE when the priority queue is full.

\begin{table}[t]
\centering
\caption{\small{Resource usage and path delay comparison where  HEFT\textsubscript{RT1} is based on $P=16$, $D=132$, $W(Avg_{TID})=16$, and HEFT\textsubscript{RT2} is based on $P=4$, $D=64$, $W(Avg_{TID})=32$. \cite{hrhs} is based on 16 PEs with a total queue length of 132. \cite{fpga_edf} is based on 1 PE with a priority queue length of 64 and a priority bit-width of 32.}}
\label{tab:comparison}
\resizebox{0.8\linewidth}{!}
{
\renewcommand{\arraystretch}{1.5}
\begin{tabular}{l|c|c||c|c|}
\cline{2-5}
                                        & HEFT\textsubscript{RT1} & \cite{hrhs} & HEFT\textsubscript{RT2} & \cite{fpga_edf}\\ \hline
\multicolumn{1}{|l|}{LUTs}             & 7,598 & 4,834 & 4,360 & 2,170 \\ \hline
\multicolumn{1}{|l|}{LUT-RAM}            & 1,920 &   N/A &   160 &   N/A \\ \hline
\multicolumn{1}{|l|}{Register}          & 6,430 & 3,968 & 3,590 & 1,158 \\ \hline
\multicolumn{1}{|l|}{Path Delay (ns)}    &  5.91 & 5.001 & 3.035 & 6.122 \\ \hline
\end{tabular}
}
\vspace{-0.3cm}
\end{table}

The \emph{Priority Queue} allows insertion and removal of tasks through the front of the queue, which prevents the depth of the queue from impacting the number of cycles needed to make a scheduling decision. Sorting the queue in odd-even phases allows each cell to only depend on its immediate neighbor, which makes the path delay independent of the queue depth. Therefore, the odd-even transposition based sorting prevents the queue depth from impacting the critical path. 
Due to the minimum tree structure of the EFT Selector, the number of PEs impacts the critical path delay, while the queue depth and bit width parameters affect the LUT and register usage. The \emph{EFT Selector} and \emph{PE Handlers} take relatively fewer resources and the \emph{PE Handler} is on the critical path due to the feedback loop that goes through the adder within the \emph{PE Handler}, comparator and multiplexers in the \emph{EFT Selector}, decoder, and finally back to the \emph{PE Handler}.
Since the \emph{EFT Selector} inputs are fed through a comparison tree driven by the number of PEs, the \emph{EFT Selector} contributes to the path delay but its contribution to critical path scales logarithmically with the PE count.

In Table~\ref{tab:comparison}, we show resource usage and path delay reported by the EDF based schedulers (~\cite{hrhs,fpga_edf}) that are implemented with priority queue lengths of 132 and 64 respectively. We synthesize two versions of HEFT\textsubscript{RT} to match the two priority queue lengths and include in the table as HEFT\textsubscript{RT1} and HEFT\textsubscript{RT2}. Since bit-width of 32 is used by \cite{fpga_edf} for priority, we set $W(Avg_{TID})=32$ for HEFT\textsubscript{RT2}.
The other implementations synthesize schedulers with search~\cite{fpga_edf} and heap-based~\cite{hrhs} insertion queues that require less comparison logic compared to our shift register based queue implementation that on the other hand offers ability to insert tasks at every cycle.

The scheduler is capable of processing a ready queue of size $n$ in $3n + 3$ cycles, with a turnaround of $2n + 3$ cycles for the task to PE mapping decision of the first entry in the sorted \emph{Priority Queue}. The \emph{Priority Queue} can be filled at a rate of one task per cycle, meaning it will take $n$ cycles to fill the priority queue with the tasks that are received from the ready queue. The odd-even sorting implemented by the shift register queue takes $n$ cycles to sort for the worst case\cite{odd_even_sort}. Additionally, both sorting phases will need to happen with no shifts for the queue to be marked as sorted, resulting with additional two cycles. The priority queue can shift out tasks at a rate of one cycle per task while reading the execution times from the LUT-RAM. After dequeuing and reading from LUT-RAM, it takes the \emph{PE Handlers} and \emph{EFT Selector} one cycle to make the PE mapping decision. The time to fill the queue, complete sorting, and select the PE accumulates to $2n + 3$ cycles to generate the first scheduling decision. Shifting out and mapping the tasks at a rate of one task per cycle adds another $n$ cycles, meaning the scheduler will take at most $3n + 3$ cycles to map a ready queue of size $n$ to each of the PEs. 
The average number of cycles it takes to make a single task to PE mapping decision is then $\frac{3n + 3}{n}$. Given that latency for the design with \emph{Priority Queue} length of 512 over 4 PE configuration is 3.048 ns, average time to make a single task to PE mapping decision is 9.144 ns excluding the data transfer overhead 
to the \emph{Priority Queue} of the scheduler.

\begin{table}[t]
\centering
\caption{\small{Hardware scheduler resource usage and path delay with respect to $P$ and $D$ on Zynq ZCU102 FPGA.}}
\label{tab:scaling}
\resizebox{1.0\linewidth}{!}
{%
\centering
\begin{tabular}{ | c | c | c | c | c | c | c | c | c | c | }
\hline
\multirow{2}{*}{$P$} & 
\multirow{2}{*}{$D$} & 
\multicolumn{2}{c|}{\multirow{2}{*}{Logic LUTs}} & 
\multicolumn{2}{c|}{\multirow{2}{*}{LUT-RAM}} &
\multicolumn{2}{c|}{\multirow{2}{*}{Registers}} & 
\multicolumn{1}{c|}{\multirow{2}{*}{\shortstack{Block \\ RAM}}} &
\multicolumn{1}{c|}{\multirow{2}{*}{\shortstack{Critical \\ Path}}} \\
& & \multicolumn{2}{c|}{ } & \multicolumn{2}{c|}{ } & \multicolumn{2}{c|}{ } & &\\

\hline
4 & 64 & 2,817 & 1.03\% & 160 & 0.11\% & 2,520 & 0.46\% & 0.5 & 3.06\\
4 & 128 & 5,190 & 1.89\% & 320 & 0.22\% & 4,159 & 0.76\% & 0.5 & 3.029  \\
4 & 256 & 9,857 & 3.60\% & 640 & 0.44\% & 7,543 & 1.38\% & 0.5 & 2.976 \\

4 & 512 & 19,603 & 7.15\% & 1,280 & 0.89\% & 14,534 & 2.65\% & 0.5 & 3.048 \\

8 & 512 & 20,471 & 7.47\% & 2,560 & 1.78\% & 15,243 & 2.78\% & 0.5 & 4.637  \\
16 & 512 & 22,038 & 8.04\% & 3,200 & 2.22\% & 16,422 & 3.00\% & 3.5 & 6.875  \\
\hline
\end{tabular}
}
\vspace{-6pt}
\end{table}

In Table~\ref{tab:scaling} we first fix the number of PEs ($P$) on the SoC to 4 and show the hardware scheduler resource usage and path delay with respect to increase in \emph{Priority Queue} length ($D$) from 64 to 512. In the same table, we fix $D$ at 512 and increase $P$ from 4 to 8 and 16. Increasing the $D$ does not affect the path delay confirming our earlier complexity analysis, and as expected LUT and LUT-RAM usage increases proportionally with $D$. As expected path delay grows less than linearly as we increase $P$ from 4 to 8 and 16 due to the minimum tree structure of the \emph{EFT Selector} unit. The EDF hardware scheduler~\cite{hrhs} that uses 16 PEs with a queue length of 134 has a critical path delay of 6.122 ns, while our design with 16 PEs and 512 queue length has a critical path delay of 6.875 ns.

\subsection{Functional Verification}

\begin{figure}[t]
    \centering
    \includegraphics[width=0.8\linewidth]{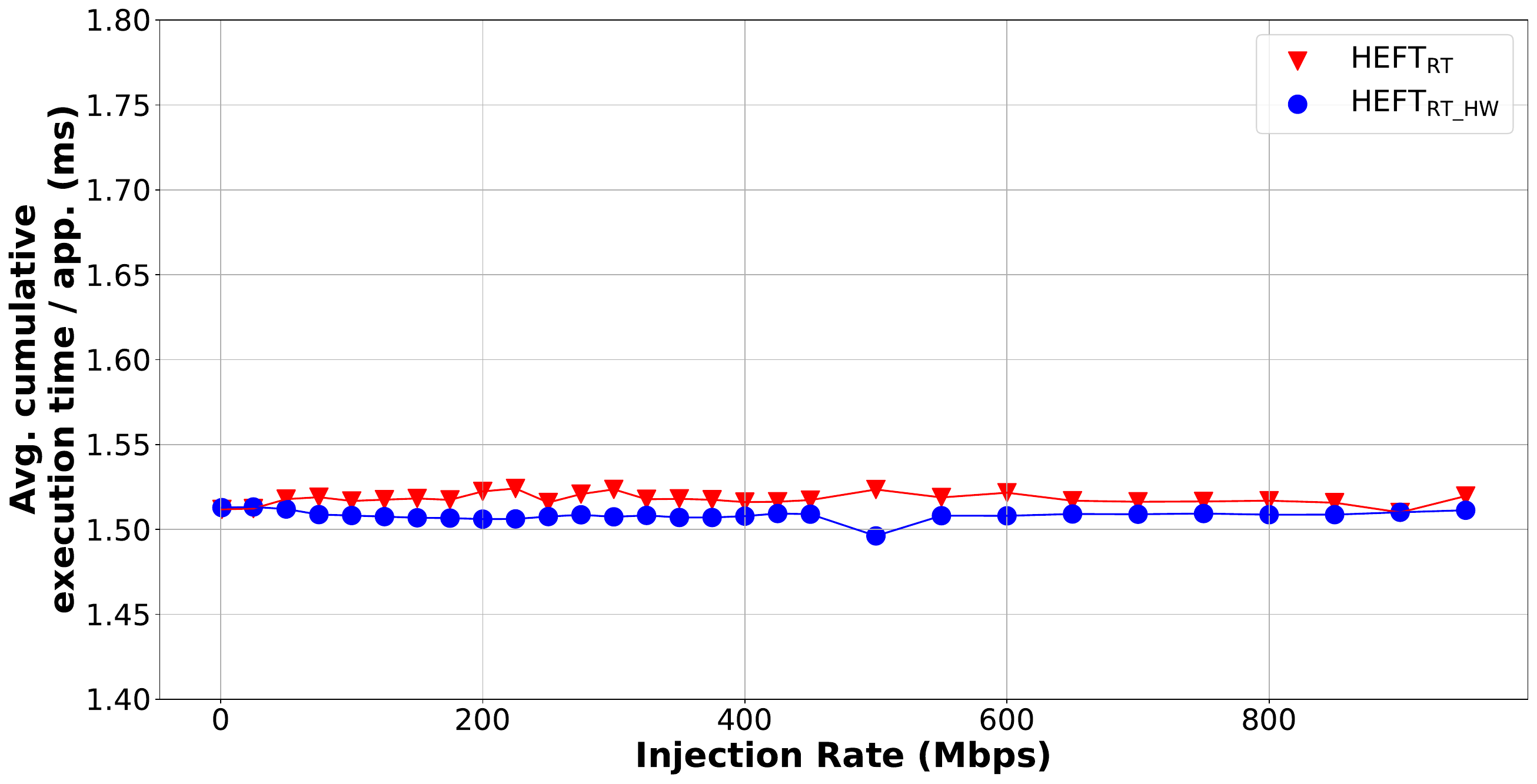}
    \caption{Minimum cumulative execution time of tasks for low latency workload versus injection rate.}
    \label{fig:cum_exec_MIN_LOW}
    \vspace{-10pt}
\end{figure}

Figure \ref{fig:cum_exec_MIN_LOW} shows cumulative execution time between software and hardware versions of the HEFT\textsubscript{RT} scheduler for low-latency workload with respect to change in job injection rate. 
We choose low-latency workloads for this scenario due to the fact that, as noted in Section~\ref{sec:setup}, they do not heavily stress the scheduler. 
The cumulative execution time is independent from scheduling overhead, and is completely dependent on task execution time. 
If the hardware scheduler were making different PE mapping decisions, we would expect to see a difference in the cumulative execution time, however the average difference across injection rates between the two schedulers is only 0.024 ms which corresponds to 0.32\% difference on average between pairwise points on the plot. This negligible difference confirms that the two schedulers are making similar task to PE mapping decisions across all injection rates. 

\begin{figure}[t]
    \centering
    \includegraphics[width=0.8\linewidth]{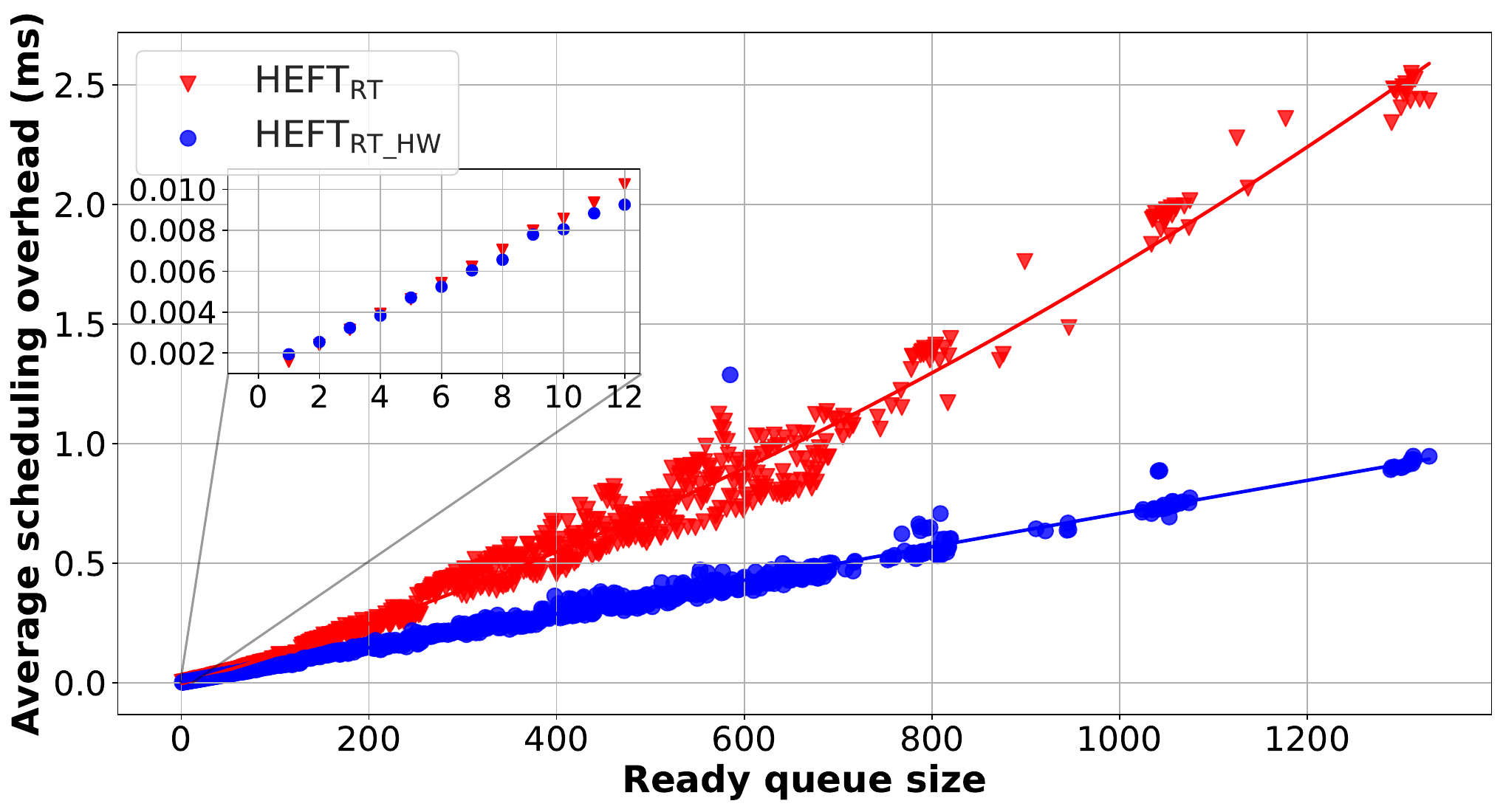}
    \caption{Ready queue size versus scheduling overhead for HEFT\textsubscript{RT} and HEFT\textsubscript{RT\_HW}}
    \label{fig:rq_vs_schedoverhead}
    \vspace{-6pt}
\end{figure}

\begin{figure}[t]
    \centering
    \includegraphics[width=0.8\linewidth]{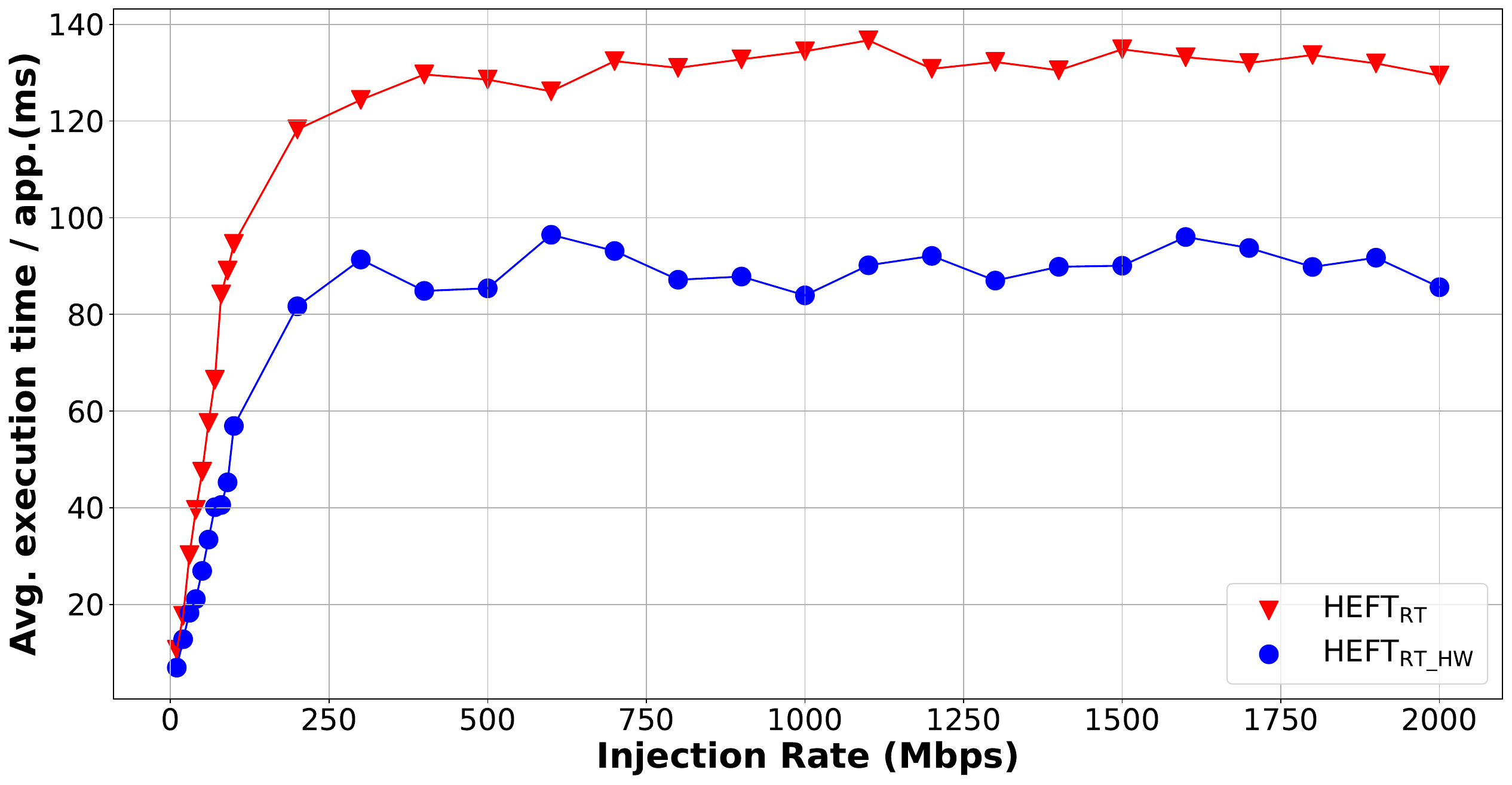}
    \caption{Average execution time vs. injection rate for HEFT\textsubscript{RT} and HEFT\textsubscript{RT\_HW} schedulers over high-latency workload.}
    \label{fig:exectime_analysis}
\vspace{-10pt}
\end{figure}

\subsection{Performance Analysis}
In Figure~\ref{fig:rq_vs_schedoverhead} we show the average scheduling overhead for software (HEFT\textsubscript{RT}) and  hardware (HEFT\textsubscript{RT\_HW}) based implementations with respect to increase in the ready queue size of the CEDR. We measure the scheduling overhead by recording the start and completion times of each mapping event. The rate of increase in scheduling overhead as the queue size grows is higher with the HEFT\textsubscript{RT} compared to the HEFT\textsubscript{RT\_HW}. 
This is expected based on the $3n + 3$ cycle count analysis presented earlier, that offers $O(n)$ complexity, while the software implementation relies on sorting with $O(n\log_{2}n)$ complexity. However, we note that in the plot we observe some outlier points that are not following the expected trend line for the HEFT\textsubscript{RT\_HW} primarily caused by the data transfer overhead on the Zynq ZCU102. In the same plot we show the ready queue size in the range of 1 to 10 in magnified form to illustrate the crossover point where HEFT\textsubscript{RT} offers less overhead for queue size of up to 5. Beyond this point, data transfer overhead to the HEFT\textsubscript{RT\_HW} is compensated by its ability to make faster scheduling decisions. The gap between the two schedulers grow as the queue size increases, where we observe a speed up of 2.6$\times$ at queue size of 1330. Without the data transfer overhead, in terms of time to make scheduling decisions HEFT\textsubscript{RT\_HW} achieves a speedup of 183$\times$.

\begin{figure}[t]
    \centering
    \includegraphics[width=0.8\linewidth]{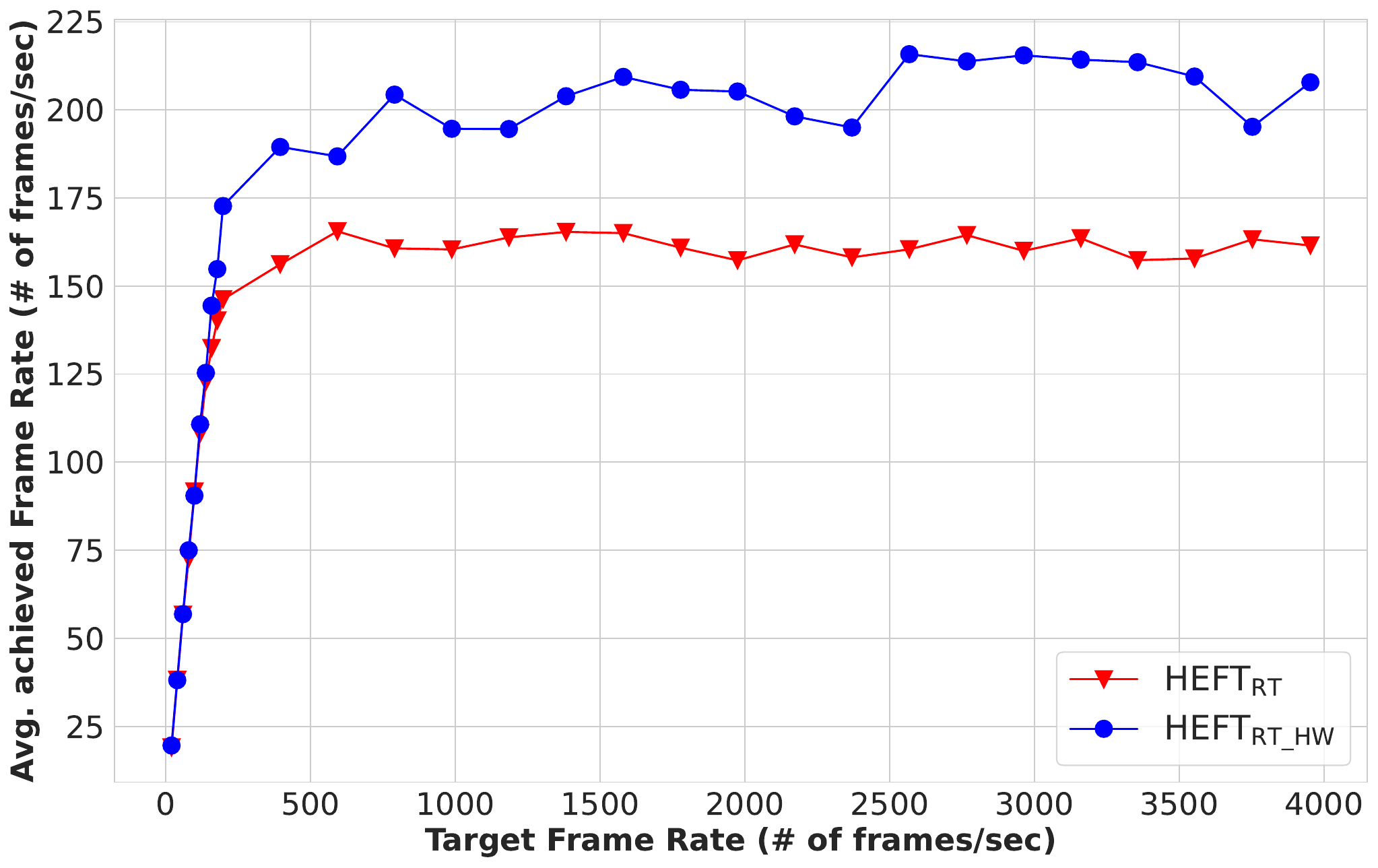}
    \caption{Target frame rate vs. average achieved frame rate for HEFT\textsubscript{RT} and HEFT\textsubscript{RT\_HW} over high-latency workload.}
    \label{fig:framerate_analysis}
\vspace{-10pt}
\end{figure}

Figure~\ref{fig:exectime_analysis} shows the average application execution time per application along the y-axis, with respect to varying injection rates along the x-axis, for both HEFT\textsubscript{RT} and HEFT\textsubscript{RT\_HW} schedulers. Here, we observe that for lower injection rates ($<$ 250 Mbps), the application execution times for both schedulers increase linearly with increasing injection rates. This is expected as increasing injection rate grows the number of concurrent jobs in the runtime, and with these jobs sharing the same set of PEs, their execution times see a linear growth. The reduction in execution time per application with the hardware scheduler becomes more dominant as the injection rate increases. As the injection rate goes beyond a certain level ($>$ 250 Mbps), the execution time per application starts to saturate on average, at 131.37 ms and 89.79 ms for HEFT\textsubscript{RT} and HEFT\textsubscript{RT\_HW} respectively. The saturation is caused by the runtime reaching to an oversubscribed state, where it is processing tasks on PEs at the maximum possible rate, and no further improvement is attainable. Execution time comparison between both schedulers across all injection rates show that with the HEFT\textsubscript{RT\_HW} this time is consistently lower than HEFT\textsubscript{RT} and results with a 31.7\% reduction on average in the saturated region. 
The improvement in execution time is achieved due to the ability of HEFT\textsubscript{RT\_HW} in making faster scheduling decisions, which enhances the runtime's capability to cope with scenarios where larger number of applications may arrive.
We further discuss this frame processing capability using  Figure~\ref{fig:framerate_analysis}, where we present 
the trend in average achieved frame rate on y-axis with respect to target frame rate on x-axis for \textit{high} latency workload.

In Figure~\ref{fig:framerate_analysis}, for lower values of target frame rates ($<$250 frames/second), both the HEFT\textsubscript{RT} and HEFT\textsubscript{RT\_HW} schedulers show similar linear trend of growth in average achieved frame rate, as the target frame rate increases. For higher values of target frame rates ($>$250 frames/second), the average achieved frame rate saturates for both the schedulers, which indicates that the runtime has become oversubscribed and system resources are fully occupied. The saturation point on the achieved frame rate with HEFT\textsubscript{RT\_HW} scheduler is at 204.62 frames/second (on average), while HEFT\textsubscript{RT} saturates at 161.51 frames/second on average. We notice that with the HEFT\textsubscript{RT\_HW} scheduler, the runtime is able to achieve saturation at around 43.11 frames/second higher frame rate on average compared to the HEFT\textsubscript{RT} scheduler. The ability of the hardware scheduler to make task-to-PE mapping decisions faster, provides the runtime with a better capability to process frequently arriving jobs at a higher rate in the oversubscribed state and in turn results with 26.7\% more frames/second in average achieved frame rate.

\section{Conclusion}\label{sec:conclusions}

In this work, we explore the design, implementation, and evaluation of the  HEFT\textsubscript{RT}~\cite{Mack22} in hardware to accelerate the scheduling of dynamically interleaved applications across heterogeneous SoCs.
We implement HEFT\textsubscript{RT} as an overlay processor on the Zynq Ultrascale+ ZCU102 FPGA and integrated it into an open source runtime system (CEDR~\cite{MackTECS22}).
We find that, for an SoC composed of 3 ARM cores and 1 FFT accelerator emulated on the same ZCU102 FPGA, we are able to generate competitive schedules with a 9.144 ns average latency for making a task-to-PE mapping decision.
We present analysis of this hardware design by exploring the impact of queue size and number of processing elements on the latency and resource usage, and we find that our design is overall quite scalable to different system configurations.
We then stress the scheduler by evaluating it in our runtime environment with realistic workloads, and we find that it can simultaneously produce high quality scheduling decisions and maintain correct application execution.
As future work, we will explore acceleration of energy-aware scheduling heuristics in order to expand our evaluations beyond focusing purely on optimization of execution time.

\bibliographystyle{IEEEtran}
\bibliography{main.bib}

\begin{thebibliography}{10}
\providecommand{\url}[1]{#1}
\csname url@samestyle\endcsname
\providecommand{\newblock}{\relax}
\providecommand{\bibinfo}[2]{#2}
\providecommand{\BIBentrySTDinterwordspacing}{\spaceskip=0pt\relax}
\providecommand{\BIBentryALTinterwordstretchfactor}{4}
\providecommand{\BIBentryALTinterwordspacing}{\spaceskip=\fontdimen2\font plus
\BIBentryALTinterwordstretchfactor\fontdimen3\font minus
  \fontdimen4\font\relax}
\providecommand{\BIBforeignlanguage}[2]{{%
\expandafter\ifx\csname l@#1\endcsname\relax
\typeout{** WARNING: IEEEtran.bst: No hyphenation pattern has been}%
\typeout{** loaded for the language `#1'. Using the pattern for}%
\typeout{** the default language instead.}%
\else
\language=\csname l@#1\endcsname
\fi
#2}}
\providecommand{\BIBdecl}{\relax}
\BIBdecl

\bibitem{cdsc}
J.~Cong, V.~Sarkar, G.~Reinman, and A.~Bui, ``{Customizable Domain-Specific
  Computing},'' \emph{IEEE Design \& Test of Comput.}, vol.~28, no.~2, pp.
  6--15, 2011.

\bibitem{Mack22}
J.~Mack, S.~E. Arda, U.~Y. Ogras, and A.~Akoglu, ``{Performant, Multi-Objective
  Scheduling of Highly Interleaved Task Graphs on Heterogeneous System on Chip
  Devices},'' \emph{IEEE Trans. on Parallel and Distrib. Syst.}, vol.~33,
  no.~9, pp. 2148--2162, 2022.

\bibitem{Topcuoglu02}
H.~{Topcuoglu} and S.~{Hariri}, ``Performance-effective and low-complexity task
  scheduling for heterogeneous computing,'' \emph{IEEE Trans. on Parallel and
  Distrib. Syst.}, vol.~13, no.~3, pp. 260--274, 2002.

\bibitem{fpga_edf}
Y.~Tang and N.~W. Bergmann, ``{A Hardware Scheduler Based on Task Queues for
  FPGA-Based Embedded Real-Time Systems},'' \emph{IEEE Trans. on Comput.},
  vol.~64, no.~5, pp. 1254--1267, 2015.

\bibitem{hrhs}
D.~Derafshi, A.~Norollah, M.~Khosroanjam, and H.~Beitollahi, ``{HRHS: A
  High-Performance Real-Time Hardware Scheduler},'' \emph{IEEE Trans. Parallel
  Distrib. Syst.}, vol.~31, no.~4, pp. 897--908, 2020.

\bibitem{hw_pfair}
N.~Gupta, S.~K. Mandal, J.~Malave, A.~Mandal, and R.~N. Mahapatra, ``{A
  Hardware Scheduler for Real Time Multiprocessor System on Chip},'' in
  \emph{2010 23rd Int. Conf. on VLSI Design}, 2010, pp. 264--269.

\bibitem{red}
L.~Kohútka and V.~Stopjaková, ``{ASIC Architecture and Implementation of RED
  Scheduler for Mixed-Criticality Real-Time Systems},'' in \emph{2020 27th Int.
  Conf. on Mixed Design of Integr. Circuits and Syst. (MIXDES)}, 2020, pp.
  83--88.

\bibitem{rt_scheduling}
G.~C. Buttazzo, \emph{{Hard Real-Time Computing Systems: Predictable Scheduling
  Algorithms and Applications}}.\hskip 1em plus 0.5em minus 0.4em\relax
  Springer Science \& Business Media, 2011, vol.~24.

\bibitem{pfair}
J.~Goossens, S.~Funk, and S.~Baruah, ``{Priority-Driven Scheduling of Periodic
  Task Systems on Multiprocessors},'' \emph{Real-Time Syst.}, vol.~25, no.~2,
  pp. 187--205, 2003.

\bibitem{hw_heft}
I.~Aliyev, J.~Mack, N.~Kumbhare, A.~Akoglu, and H.~F. Ugurdag, ``{FPGA-based
  Minimal Latency HEFT Scheduler for Heterogeneous Computing},'' in \emph{2021
  6th Int. Conf. on Comput. Sci. and Eng. (UBMK)}, 2021, pp. 1--5.

\bibitem{MackTECS22}
J.~Mack, S.~Hassan, N.~Kumbhare, M.~Gonzales, and A.~Akoglu, ``{CEDR - A
  Compiler-integrated, Extensible DSSoC Runtime},'' \emph{ACM Trans. on
  Embedded Comput. Syst.}, vol.~33, no.~9, pp. 2148--2162, 2022.

\bibitem{maurya2018benchmarking}
A.~K. Maurya and A.~K. Tripathi, ``On benchmarking task scheduling algorithms
  for heterogeneous computing systems,'' \emph{The J. of Supercomputing},
  vol.~74, no.~7, pp. 3039--3070, 2018.

\bibitem{odd_even_sort}
D.~Bitton, D.~J. DeWitt, D.~K. Hsaio, and J.~Menon, ``{A Taxonomy of Parallel
  Sorting},'' \emph{ACM Comput. Surv.}, vol.~16, no.~3, p. 287–318, 1984.

\end{thebibliography}

\end{document}